\documentclass[fp,twocolumn]{jpsj3}
\newcommand{\byto}{Ba$_{3-x}$Yb$_x$Ta$_5$O$_{15}$}
\usepackage{txfonts}
\usepackage[T1]{fontenc}
\usepackage[utf8]{inputenc}
\usepackage{color}
\usepackage{xcolor}
\usepackage{ulem}
\usepackage{graphicx}
\usepackage{gensymb}
\begin{document}

\title{Yb $4f$-Ta $5d$ hybridization and valence evolution in tetragonal tungsten bronze Ba$_{3-x}$Yb$_x$Ta$_5$O$_{15}$}
% Force line breaks with \\

\author{Daisuke Takegami$^{1,2}$, Haruki Takei$^{3}$, Masato Yoshimura$^{4}$, Takuro Katsufuji$^{3}$, Takashi Mizokawa$^{1}$}
\inst{$^{1}$Department of Applied Physics, Waseda University, Shinjuku, Tokyo 169-8555, Japan\\
$^{2}$Max Planck Institute for Chemical Physics of Solids, N{\"o}thnitzer Stra{\ss}e 40, 01187 Dresden, Germany\\
$^{3}$Department of Physics, Waseda University, Shinjuku, Tokyo 169-8555, Japan\\
$^{4}$National Synchrotron Radiation Research Center, 30076 Hsinchu, Taiwan
}

%daisuke.takegami@cpfs.mpg.de
%haruki.takei@ruri.waseda.jp
%yoshimur@spring8.or.jp
%katsuf@waseda.jp
%mizokawa@waseda.jp

\date{\today}

\abst{
Here we investigate the electronic structure of the  tetragonal tungsten bronze Ba$_{3-x}$Yb$_x$Ta$_{5}$O$_{15}$ by making use of hard x-ray photoemission spectroscopy. The core level spectroscopy shows that the substitution with Yb ions in the series first occurs on the compact S1 site. For $x\leq1$, Yb is found to be dominantly Yb$^{2+}$ with a small mixing of Yb$^{3+}$, while for $x>1$, a significant increase of Yb$^{3+}$ is found, suggesting not only that site S2 favours Yb$^{3+}$, but also that their presence affects also the valency of the ions in site S1. The valence band spectra shows a relatively deep Yb$^{2+}$ doublet, but at the same time indications of a Ta~$5d$-Yb~$4f$ interaction are found, suggesting the presence of Yb~$4f$ carriers at the Fermi level through this hybridization. Our results thus point towards an exotic form of $d$-$f$ electronic interplay that together with the structural degrees of freedom can result in the unusual trends observed in the physical properties of Ba$_{3-x}$Yb$_x$Ta$_{5}$O$_{15}$.}

\maketitle

\section{Introduction}
Transition-metal oxides exhibit a variety of interesting physical properties due to strong electron-electron interaction and electron-lattice interaction \cite{Imada98}. The interplay between the electron-electron and electron-lattice couplings has been extensively studied in 3$d$ transition-metal oxides. For example, orbital orderings through Kugel-Khomskii mechanism are derived from the electron-electron interaction while those through Jahn-Teller mechanism rely on the electron-lattice interaction on localized and degenerate electronic systems \cite{Khomskii2014}. Also transition-metal oxides such as VO$_2$ and NbO$_2$ exhibit insulating ground state induced by metal-metal dimerization through strong electron-lattice interaction \cite{Hiroi2015}. Compared to the 3$d$ transition-metal oxides, the relationship between the electron-electron and electron-lattice couplings is not well explored in 4$d$ and 5$d$ transition-metal oxides. Since the hybridization with the O 2$p$ orbitals is stronger for the transition-metal 4$d$ or 5$d$ orbitals than that for the transition-metal 3$d$ orbitals, the effect of electron-lattice coupling is expected to be more spectacular in the 4$d$ or 5$d$ systems.
The oxides with tetragonal tungsten bronze structure is one of such interesting systems. Ba$_{3-x}$Sr$_x$Nb$_{5}$O$_{15}$ exhibits a metal-to-insulator transition by the Sr doping. While Ba$_{3-x}$Sr$_x$Nb$_{5}$O$_{15}$ is a good metal with partially filled Nb 4$d$ band for $0 \leq x \leq 2$, it shows semiconducting or insulating behavior at low temperature for $x \geq 3$ \cite{Kolodiazhnyi2015}. As shown in Fig. 1, the NbO$_6$ octahedra share their corners and form a three dimensional network \cite{Jamieson1968,Hessen1991a,Hessen1991b,Hwang1997}. The Ba ions are surrounded by the NbO$_6$ octahedra in the two different ways. The compact S1 site surrounded by eight NbO$_6$ octahedra has 12 nearest oxygen atoms \cite{Jamieson1968}. The spacious S2 site is surrounded by ten NbO$_6$ octahedra and six of them are rather close to the Nb ion than the others. The tetragonal or orthorhombic unite cell includes two S1 sites and four S2 sites. It has been pointed out that the smaller (larger) ions tend to occupy the S1 (S2) site \cite{Jamieson1968}. The evolution of the electronic structure by Sr substitution for Ba has been studied by transport \cite{Kondoh2021} and photoemission spectroscopy
\cite{Yasuda2020,Nakamura2022}.
In addition, the interplay between the magnetic rare-earth elements and the Nb 4$d$ conducting electrons has been elucidated \cite{Iwamoto2022,Sekino2023,Nakamura2024_ENO} 

Compared to the Nb systems, its 5$d$ counterpart Ba$_3$Ta$_{5}$O$_{15}$ and its relatives are less explored. They exhibit the same tetragonal tungsten bronze structure  
\cite{Feger1995,Siegrist1997,Kim2015,Du2021}.
Very recently, Takei et al. have studied the interplay between the magnetic rare-earth element $R$ and the Ta~5$d$ conducting electrons in Ba$_{3-x}R_x$Ta$_{5}$O$_{15}$
\cite{Takei2024}. Among them, Ba$_{3-x}$Yb$_x$Ta$_{5}$O$_{15}$ exhibits interesting Yb valence change which would be related to the unique lattice properties and the spin/charge/orbital degrees of freedom of {the Yb 4$f$/Ta 5$d$ states}~\cite{Takei2024_Yb}. 
{In their study, Takei et al. find that the amount of magnetic Yb$^{3+}$ ions expected from the increasing bulk magnetic susceptibility with x, outpaces that of the concentration x, suggesting a population largely consisting of Yb$^{2+}$ for small x, with a gradual replacement of the Yb$^{2+}$ ions to Yb$^{3+}$ with increasing x. At the same time, however, starting from a highly metallic behaviour at x=0, the resistivity shows an increasing trend with x, and together with the Hall coefficients, they indicate that the number of the Ta~5$d$ conduction electrons is expected to decrease with x. This shows an apparent contradiction, as the increase of the mean Yb valence as observed in the susceptibility should, in principle, result in a lower average Ta valence, resulting in more Ta~5$d$ carriers.}
In the present work, the electronic structure of Ba$_{3-x}$Yb$_x$Ta$_{5}$O$_{15}$ ($x$=0.0, 0.5, 1.0, and 1.2) is examined by means of hard x-ray photoemission spectroscopy to provide insights of {this} interesting interplay between the Yb valence and the Ta 5$d$ electrons under the unique tetragonal tungsten bronze lattice.

\begin{figure}[h]
\begin{center}
\includegraphics[width=0.96\columnwidth]{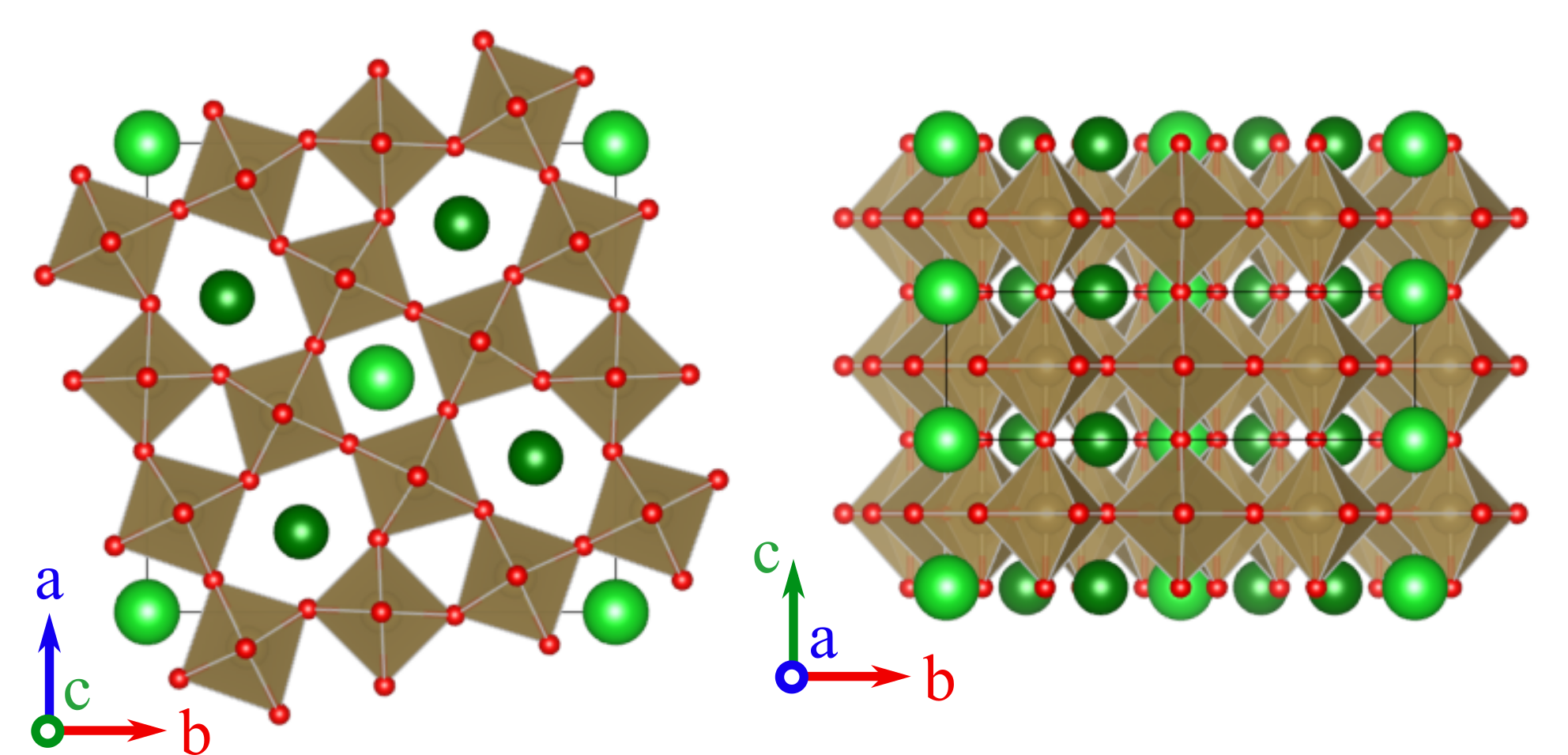}
\end{center}
\caption{Crystal structure of \byto\ illustrated with the program VESTA~\cite{VESTA}. The brown octahedra correspond to the TaO$_6$, the red spheres to the O ions and the green and dark green spheres to the two distinct Ba sites S1 and S2.}
\label{Fig_struct}
\end{figure}

\section{Methods}
Single crystal samples of \byto\ were synthesized as reported in the literature~\cite{Takei2024_Yb}. Details on the synthesis process, structural characterization and bulk properties of the measured samples have been reported in the previous work~\cite{Takei2024_Yb}. Hard x-ray photoelectron spectroscopy (HAXPES) measurements were performed at the Max-Planck-NSRRC endstation at the Taiwan BL12XU beamline of SPring-8~\cite{Weinen2015,Takegami2019}. The crystals were cleaved under ultrahigh vacuum in order to expose a clean surface and the measurements were performed at 300~K. The pressure was in the low $10^{-10}$~mbar range in the spectrometer chamber and in the $10^{-9}$~mbar range in the cleaving chamber. The photon energy was set to 6.5~keV and the total energy resolution was about 250~meV. The beam is linearly polarized, with an MB Scientific A-1 HE analyser located at the direction parallel to the polarization. The samples were oriented with the c axis parallel to the polarization direction.  Local-density approximations (LDA) calculations were performed using the full-potential local-orbital code FPLO~21~\cite{Koepernik99}.

\section{Results}

\begin{figure}[h]
\begin{center}
\includegraphics[width=0.98\columnwidth]{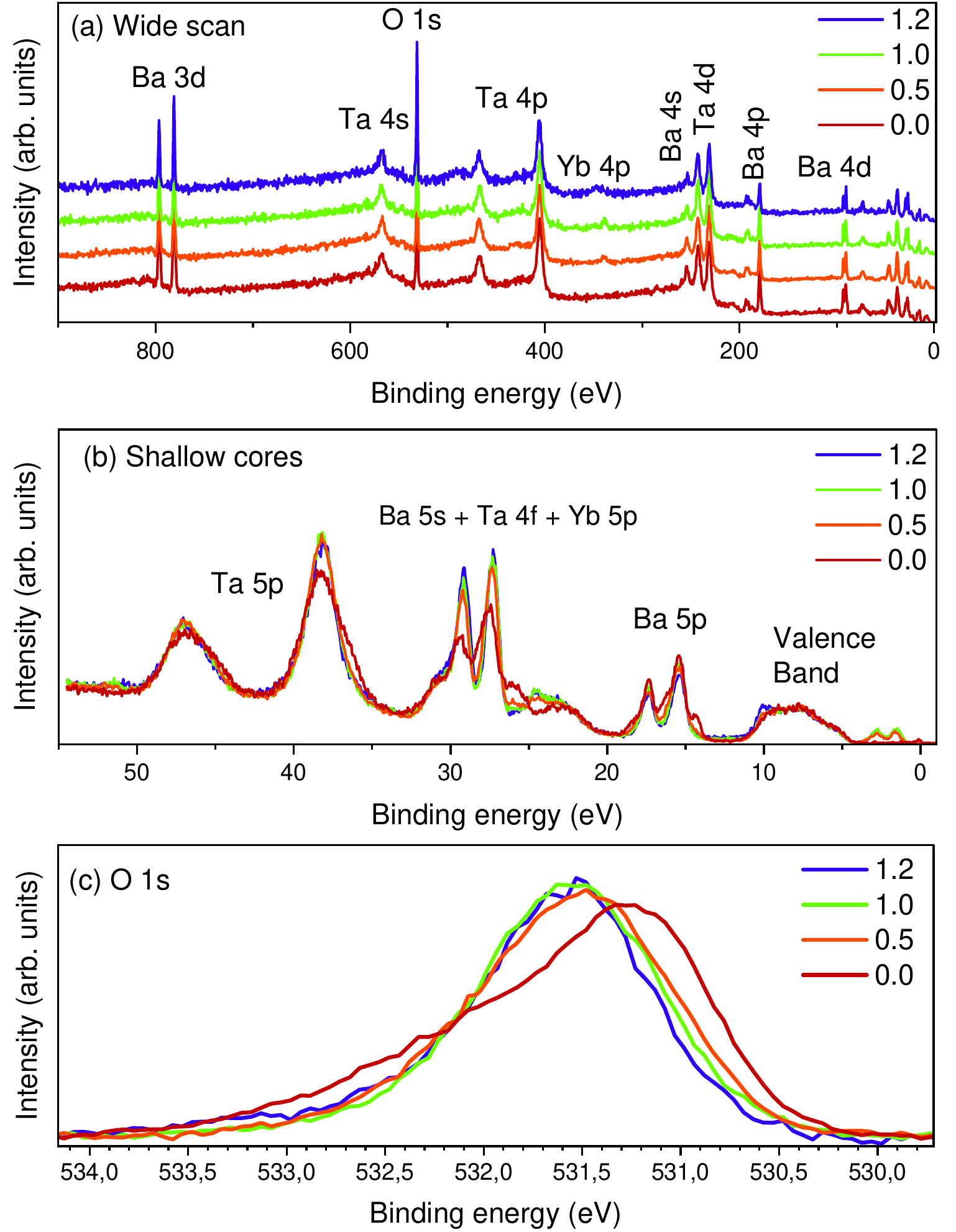}
\end{center}
\caption{HAXPES wide scan of \byto\ (a), finer close-up scans below 55~eV (b), and the O~1s core level {spectra} (c).}
\label{Fig_overview}
\end{figure}

Figure~\ref{Fig_overview}(a) shows the wide scan overview of \byto. All features observed in Fig.~\ref{Fig_overview}(a) correspond to core levels of Ba, Ta, O, and Yb (for $x\neq0$), indicating the lack of contaminants and confirming the cleanliness of the cleaved surface. All spectra in Fig.~\ref{Fig_overview} have been normalized to the integrated intensity of the O~$1s$ core level {spectra}. Several changes not only in intensity according to the composition, but also in the peak shapes depending on the composition can be observed in Fig.~\ref{Fig_overview}(b), as we will discuss later in the text.

Figure~\ref{Fig_overview}(c), shows the O~$1s$ {spectra}. Here, we observe that the $x=0$ displays a strongly asymmetric main peak line in contrast to the more symmetric lines for $x\neq0$. This suggests a stronger metallicity for $x=0$, with in particular, the participation of the O states at the Fermi level. Possible O site disorders due to the large Ba ions confined in the compact S1 site might also contribute to a high-energy component that results in this strong asymmetry, as also observed for Ba$_3$Nb$_5$O$_{15}$\cite{Yasuda2020}.
As for $x=0.5$ and $x=1.0$, the lower energy side of the main peak extends by $100~$meV lower than in $x=1.2$, suggesting a mild lower energy component from metallic screenings not present in $x=1.2$. Here we note that the higher energy side of the $x\neq0$ peaks coincide very well, indicating that it is not a rigid shift of the peaks.

\begin{figure}[h]
\begin{center}
\includegraphics[width=0.98\columnwidth]{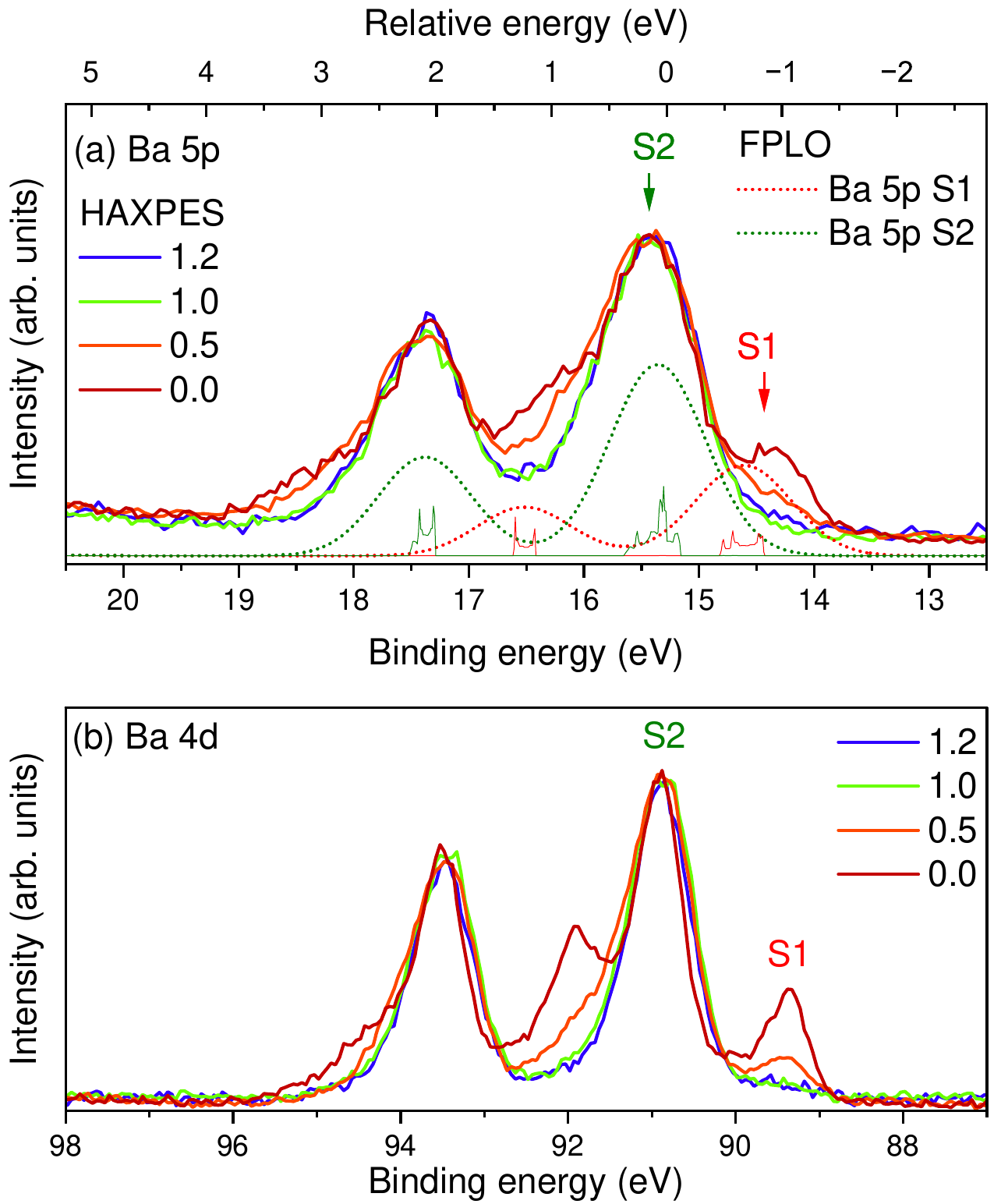}
\end{center}
\caption{(a) Ba~$5p$ core level {HAXPES spectra} of \byto, together with the calculated Ba~$5p$ PDOS in Ba$_3$Ta$_5$O$_{15}$. (b) Ba~$4d$ core level {HAXPES spectra} of \byto. The intensity has been normalized to the maximum of the peak labelled S2 in order to allow for a better lineshape comparison.}
\label{Fig_Bacores}
\end{figure}

Figure~\ref{Fig_Bacores}(a), shows the Ba~$5p$ core level {HAXPES spectra}, which is split by the spin-orbit coupling into Ba~$5p_{1/2}$ and Ba~$5p_{3/2}$ {contributions}, with a splitting of around 2~eV. For $x=0$, and up to a lesser extent, $x=0.5$ we note that there is an additional component approximately 1~eV lower to the main peaks. This is more clearly observed in the intrinsically sharper Ba~$4d$ core level {lineshape}, shown in Fig.~\ref{Fig_Bacores}(b). These two distinct components in $x=0$ arise from the two distinct Ba sites in the crystal structure of Ba$_3$Ta$_5$O$_{15}$. Site~S1 has a multiplicity of 2 and is surrounded much more closely by the TaO$_6$ octahedra than that in S2 (with multiplicity of 4). This suggests a significantly different Madelung potential for the two sites, resulting in the two distinct sets of peaks in the core level {spectra} with approximately 1:2 ratio. Figure~\ref{Fig_Bacores}(a) confirms this interpretation by showing that the Ba~$5p$ partial density of states (PDOS) from the ab-initio LDA calculations reproduce this 1~eV difference in the S1 and S2 contributions. For $x>0$, we observe that the S1 peak gets quickly suppressed, disappearing entirely at $x=1$. This shows that the replacement of the Ba ions by Yb occurs first in the compact S1 site. As for the small high-energy component visible for $x=0.0$, this is once again consistent with possible O site disorders as discussed above for the O~$1s$ {spectra}, although high-resolution structural studies would be necessary to confirm such hypothesis.

\begin{figure}[h]
\begin{center}
\includegraphics[width=0.98\columnwidth]{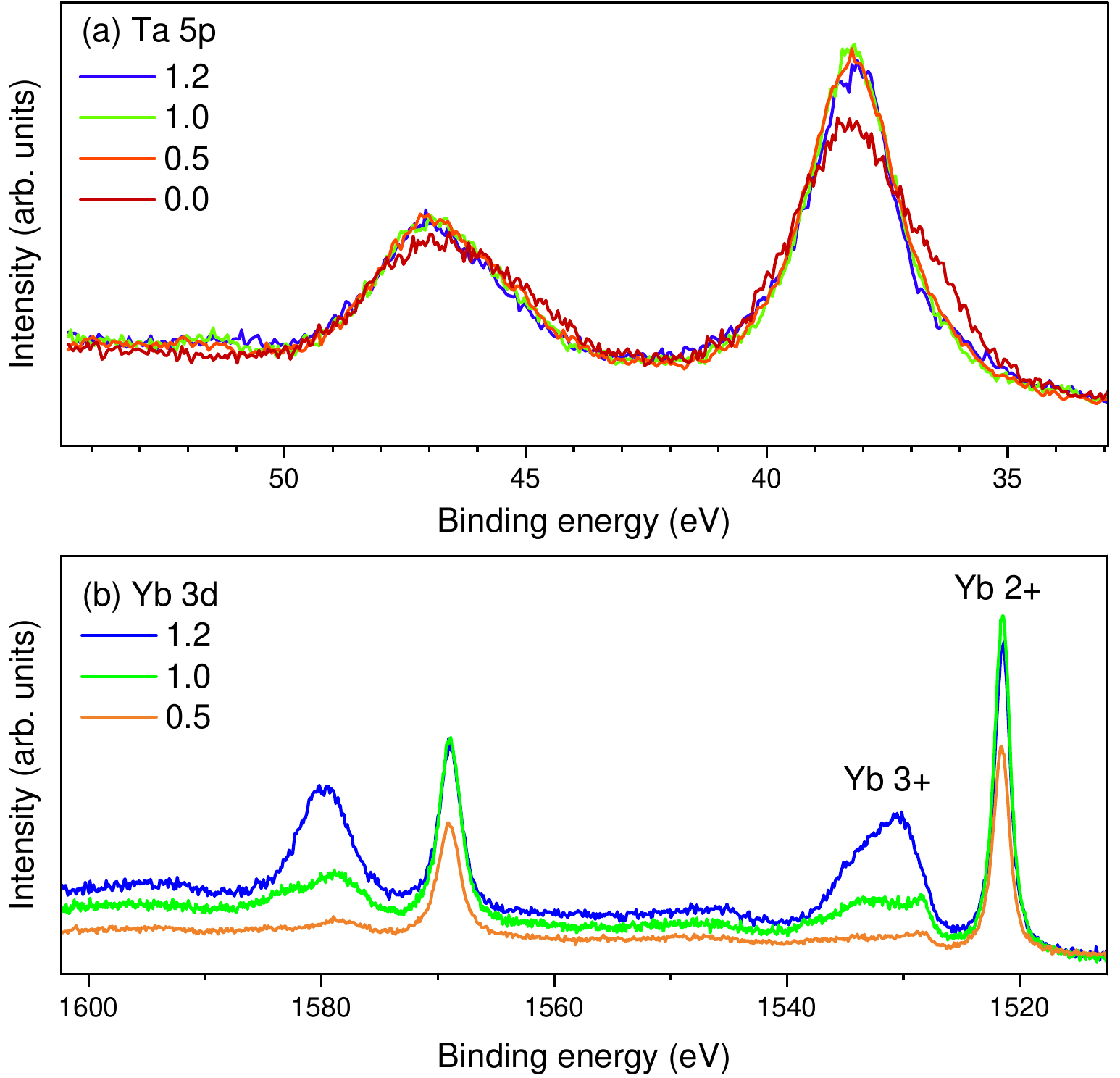}
\end{center}
\caption{(a) Ta~$5p$ and (b) Yb~$3d$ core level {HAXPES spectra} of \byto.}
\label{Fig_cores}
\end{figure}

Figure~\ref{Fig_cores}(a) shows the Ta~$5p$ core level {spectra}. For $x>0$, the peaks look symmetric and with no appreciable changes. In the case of $x=0.0$, however, we observe that there is a lower energy component taking spectral weight from the main peak. This is similar to as observed in Ba$_3$Nb$_5$O$_{15}$ and its derivative compounds, in which this lower energy component arises due to metallic screening. Interestingly, for all $x>0$, this component seems to be completely vanished, in line also with the other core levels showing less metallic-like features for $x>0$.  Here we note that the intrinsically broader Ta~$5p$ core level {lineshape} compared to {that of} the O~$1s$ does not allow for the determination of the presence of finer changes between the $x>0$ samples as it was discussed in the O~$1s$.

Figure~\ref{Fig_cores}(b) shows the Yb~$3d$ core level {spectra}.  Yb~$3d$ core level spectroscopy is well known approach to study the Yb states, due to the clear separation and distinctive multiplet shapes for the different final state valencies~\cite{Moreschini2007,Yamaguchi2009, Utsumi2012, Stavinoha2024}. Here we observe one single sharp peak at around 1520~eV corresponding to the Yb$^{2+}$ $3d_{5/2}$ final state, and broader features corresponding to the Yb$^{3+}$ multiplets around 10~eV higher in binding energy. The same is repeated 50~eV higher for the $3d_{3/2}$ contributions. Overall, we observe a rather linear increase between $x=0.5$ and $x=1.0$, suggesting that the Yb ions added for $x\leq1.0$, which occupy S1 as discussed with the Ba core level {spectra}, have a dominantly $2+$ character with a small Yb$^{3+}$ mixing. For $x=1.2$, however, the Yb$^{2+}$ peak actually decreases in intensity, and the Yb$^{3+}$ significantly increases, indicating not only that the ions included in S2 might favour a higher Yb$^{3+}$ weight, but also, that it is affecting the valence mixing of the ions in S1 towards Yb$^{3+}$.
{We note that, while Yb$^{3+}$ ions are expected to have a slightly smaller ionic radius than that of Yb$^{2+}$ ions, the Yb$^{3+}$ contributions seem to dominantly originate from the Yb in the spacious S2 site, rather than replacing those in the S1 site. This is evidenced by the lack of splitting or broadening in the Yb$^{2+}$ peak at $x=1.2$ that should have been expected due to potential differences between S1 and S2 as previously seen in the Ba~$4p$ and $5p$ spectra.}
Furthermore, we note that the Yb$^{3+}$ final state features are overall not very clearly defined, not displaying the clear distinctive individual multiplet peaks often observed. This broadening in the multiplet structures of rare earth systems is considered to be the result of a strong interaction with the ligands~\cite{Chazalviel1976,Melendez2024}, suggesting thus that the Yb$^{3+}$ states in Ba$_{3-x}$Yb$_x$Ta$_{5}$O$_{15}$ might not be fully atomic-like, but rather, hybridizing and interacting with its surrounding ligands. In particular, the broader, less defined Yb$^{3+}$ features for $x=1.2$ with small Yb ions in the larger S2 might suggest some form of rattling effect modulating the hybridization.

\begin{figure}[h]
\begin{center}
\includegraphics[width=0.98\columnwidth]{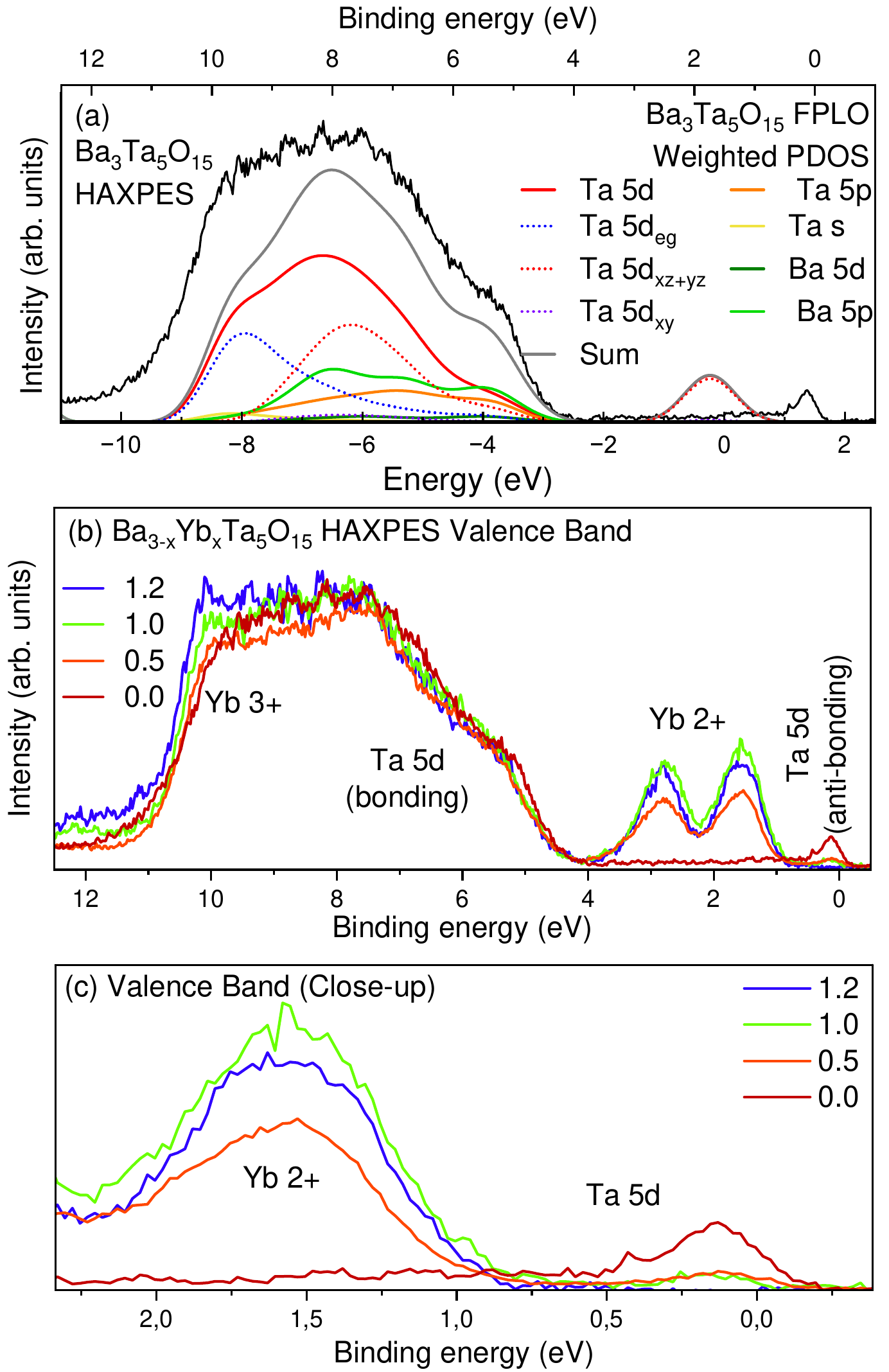}
\end{center}
\caption{
(a) HAXPES valence band spectra of Ba$_3$Ta$_5$O$_{15}$ (x=0), together with the PDOS from LDA calculations, multiplied by their respective photoionization cross-sections at $h\nu=6.5$~keV and a Gaussian broadening to account for experimental resolution.
(b) HAXPES valence band spectra of \byto\, and (c) a close-up near the Fermi level region }
\label{Fig_valence}
\end{figure}
Moving now to the valence band spectrum, we will start by showing and discussing the spectra of Ba$_3$Ta$_5$O$_{15}$ as shown in Fig.~\ref{Fig_valence}(a). Overall, in the experiment we observe a very broad set of features between 5-10~eV in binding energy, and a peak near the Fermi level with an elongated tail-like feature. In similar systems, the deeper feature has usually been associated with the O~$2p$ bands, and the feature next to the Fermi level with the transition metal d-bands. In the case of $5d$ transition metal systems, however, it is known that the HAXPES valence band generally reflects the $5d$ {contributions} due to {their} overwhelmingly large photoionization cross-sections compared to e.g. that of O~$2p$~\cite{Trzhaskovskaya2018,Takegami2020,Takegami2022_Imaging, Altendorf2023}. A simple exercise by using the occupied PDOS from LDA calculations, multiplied by its respective cross-sections shows indeed that the observed spectral features and its weights can be well understood as mainly consisting of Ta~$5d$ {contributions}, both for the deeper as well as shallower features. The deeper features consist of a mix of bonding states of Ta~$5d_{eg}$ character on the deeper site around 8-10~eV in binding energy, and Ta~$5d_{xz}$ and $5d_{yz}$ bonding states on the shallower 5-8~eV range. This spectral shape consisting of bonding and anti-bonding features is closely analogous to many other strongly covalent $5d$ systems such as ReO$_3$ or many iridates~\cite{Takegami2020,Falke2021,Takegami2022_Imaging}. We note that in the calculations there is an underestimation of the gap between the bonding and anti-bonding features because of self-interaction effects, as reported in other $4d$ and $5d$ transition metal systems~\cite{Efimenko2017,Falke2021}. A small discrepancy of the spectral weights can also be observed, in particular, with the experiment showing a more suppressed $t_{2g}$ signal and enhanced $e_{g}$ contributions. This is to be expected from the orientation dependence effects~\cite{Takegami2022_Imaging} resulting from performing the experiments on single crystals with the c axis normal to the polarization of the light, corresponding to an orbital node of $t_{2g}$ orbitals and maximum of $e_{g}$ {orbitals}. Our calculations estimate an occupation of around 0.2~$5d_{xz+yz}$ electrons per Ta in the anti-bonding bands near the Fermi level, a carrier concentration as expected from its nominal valency.

Next, we examine the changes when replacing Ba by Yb (Fig.~\ref{Fig_valence}(b)). First of all, we observe that the Yb$^{2+}$~$4f$ final state doublet shows up at around 1.5 and 3~eV, relatively deep compared to many Yb intermetallic systems where the Yb~$4f$ peak crosses the Fermi level. Similar to the core level, between $0\leq x\leq1$ the Yb$^{2+}$ feature shows a linear increase indicating that the added Yb ions are mostly $2+$, but at $x=1.2$ it actually decreases a little, indicating that there is an overall valence change that affects not only the new ions in $S2$ but also those in $S1$. This change is accompanied by an increase of spectral weight in the 8-12~eV region, that is, around 6-10~eV higher than the Yb$^{2+}$~$4f$ doublet, where the Yb$^{3+}$~$4f$ multiplets are usually expected. This increase in spectral weight does not show clear distinguishable multiplet-like features, being instead a very broad and undefined increase, similar to the Yb$^{3+}$ feature in the Yb~$3d$ core level {spectra}. This would suggest once again that the Yb$^{3+}$ states are not fully atomic-like but must be interacting with its surrounding bands, resulting in these washed out features.

Finally, we focus now on the Ta~$5d_{xz+yz}$ anti-bonding states crossing the Fermi energy. In the close-up shown in Fig.~\ref{Fig_valence}(c), we observe a clear suppression of these states by increasing $x$. This is fully consistent with the metallic screening peaks and asymmetric shapes observed only in the core levels of $x=0.0$. For $x=0.5$ and $x=1.0$, Ta~$5d_{xz+yz}$ anti-bonding feature is suppressed to almost 25\% and 20\% respectively. This becomes even more significant for $x=1.2$, where the Ta~$5d_{xz+yz}$ weight has almost disappeared, with its integrated intensity being below 10\% that of $x=0.0$.

\section{Discussions}
\begin{figure}[h]
\begin{center}
\includegraphics[width=0.98\columnwidth]{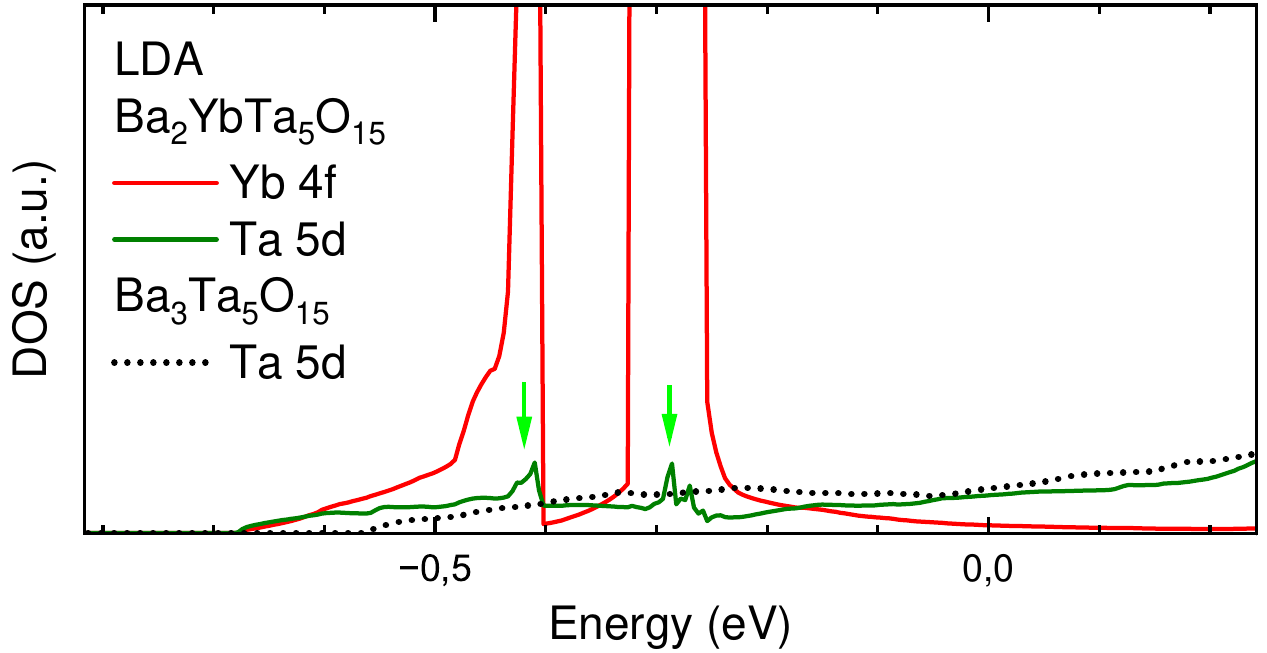}
\end{center}
\caption{Comparison of Ta~$5d$ PDOS from LDA calculations performed with FPLO for Ba$_2$YbTa$_5$O$_{15}$ and Ba$_3$Ta$_5$O$_{15}$.}
\label{Fig_BYTO_FPLO}
\end{figure}

The changes observed at the vicinity of the Fermi level suggest a change in either the concentration or the orbital character of the conduction electrons. One first assumption for such change would be to assume some sort of Fermi level shift with $x$. However, in the core level spectra no such dependencies are observed in the peak positions, indicating that such explanation is unlikely. Instead, we explore the possibility of some form of interaction between the Yb~$4f$ and the Ta~$5d_{xz+yz}$ {states}. In Fig.~\ref{Fig_BYTO_FPLO}, we show simple LDA-based comparison of Ba$_{3-x}$Yb$_x$Ta$_{5}$O$_{15}$, $x=0$ and $1$ (assuming all Yb goes to site S1). We observe that in the calculations, the Yb~$4f$ states (which are largely non-metallic, albeit shallower than in the experiment), hybridize with the Ta~$5d$ {states}, shifting DOS away from the immediate vicinity of the Fermi level and instead mixing with the deeper Yb~$4f$ {states}. While of course the simplicity of such LDA model is not enough to accurately describe the spectra nor the possible complex Yb~$4f$-Ta~$5d$ interaction, this suggests that the Ta~$5d$ states are mixed and/or altered by the presence of Yb~$4f$ {electrons} and {their} hybridization. This form of mixing would be consistent with the broader Yb features due to their interaction with surrounding bands. Furthermore, such mechanism would result in {the} Yb~$4f$ {states} partly contributing to the conduction bands through {their} mixing with the Ta~$5d$ {states}. This provides an explanation to the suggestions by Takei et al.~\cite{Takei2024} that the Yb~$4f$ electrons are expected to be contributing to the conduction, despite our observation that the Yb~$4f$ main peaks being relatively far away from the Fermi level. The photoionization cross sections per electron for Yb~$4f$ {contributions} is smaller than that {for the} Ta~$5d$ {contributions}, and thus, such mixing would result also in the decrease of the observed spectral weight.  

We thus end up in the following scenario: The Yb ions first substitute the Ba in site S1, mixed valent but dominantly Yb$^{2+}$. {This selective Yb occupation of the S1 site is consistent with the size difference between the Ba$^{2+}$ and the smaller Yb$^{2+/3+}$ ions.} The Yb~$4f$ main peaks do not directly cross the Fermi level unlike in many intermetallic Yb compounds, but our results suggest they might contribute through its mixing with the Ta~$5d$ states. By going to $x>1.0$, a sudden increase in Yb$^{3+}$ signal is observed, suggesting that the occupation of site S2 does favour the increase of Yb$^{3+}$ weight also in S1. At the same time, the spectral weight at the Fermi level is further suppressed, indicating most likely a stronger Ta~$5d$-Yb~$4f$ coupling. {The perhaps counter-intuitive occupation of the spacious S2 site with smaller Yb$^{3+}$-dominant ions, while larger Yb$^{2+}$-dominant ions remaining in the compact S1, suggests that here the effect of electronic interactions with the neighbouring ions is more important than the purely structural size differences. Determining, by other spectroscopic means, the symmetry of the Yb$^{3+}$ states might provide valuable insights on the relevant energetic considerations that are into play.}
Further theoretical studies on this from of hybridization as well as the magnetic coupling between the Yb sites and the possible states that might arise from these could {also} further expand the understanding of this phenomena. We note that possibilities such as some form of site disorder or cation deficiency suggested in previous studies~\cite{Takei2024} also might be playing a role in the carrier concentration and in general in the electronic states observed in Ba$_{3-x}$Yb$_x$Ta$_{5}$O$_{15}$, and thus, more in-depth structural studies might also be desirable.
\section{Conclusions}
In conclusion, we have characterized the electronic structure of Ba$_{3-x}$Yb$_x$Ta$_{5}$O$_{15}$ experimentally by making use of hard x-ray photoemission spectroscopy. The Ba core level spectroscopy results show two distinct sets of peaks corresponding to the Ba ions in the two different sites S1 and S2 having significantly different Madelung potentials, as validated by means of density functional theory calculations. The substitution with smaller Yb ions first occurs on the compact S1 site. For $x\leq1$, Yb is shown to be dominantly Yb$^{2+}$, although with a small mixing of Yb$^{3+}$, while at $x=1.2$, the weight of Yb$^{3+}$ significantly increases while the overall Yb$^{2+}$ weight decreases, suggesting not only that site S2 favours Yb$^{3+}$, but also that their presence affects also the valency of the ions in site S1.
In the valence band, there is a gradual suppression of the spectral weight at the Fermi level with increasing $x$, an observation that our simple calculations suggest might be caused by the spectral weight shift from {the} Ta~$5d$ to {the} Yb~$4f$ {states}. This hybridization results in the presence of Yb~$4f$ carriers also contributing to the conduction despite the relatively deep Yb$^{2+}$ doublet observed experimentally. Our results thus point towards an exotic form of $d$-$f$ electronic interplay that together with the structural degrees of freedom can result in the unusual trends observed in the physical properties of Ba$_{3-x}$Yb$_x$Ta$_{5}$O$_{15}$.
\begin{acknowledgement}
D.T. acknowledges the support by the Deutsche Forschungsgemeinschaft (DFG, German 
Research Foundation) under the Walter Benjamin Programme, Projektnummer 521584902.
We acknowledge the support for the measurements from the Max Planck-POSTECH-Hsinchu Center for Complex Phase Materials.
\end{acknowledgement}
\appendix
\section{}

\begin{figure}[h]
\begin{center}
\includegraphics[width=0.98\columnwidth]{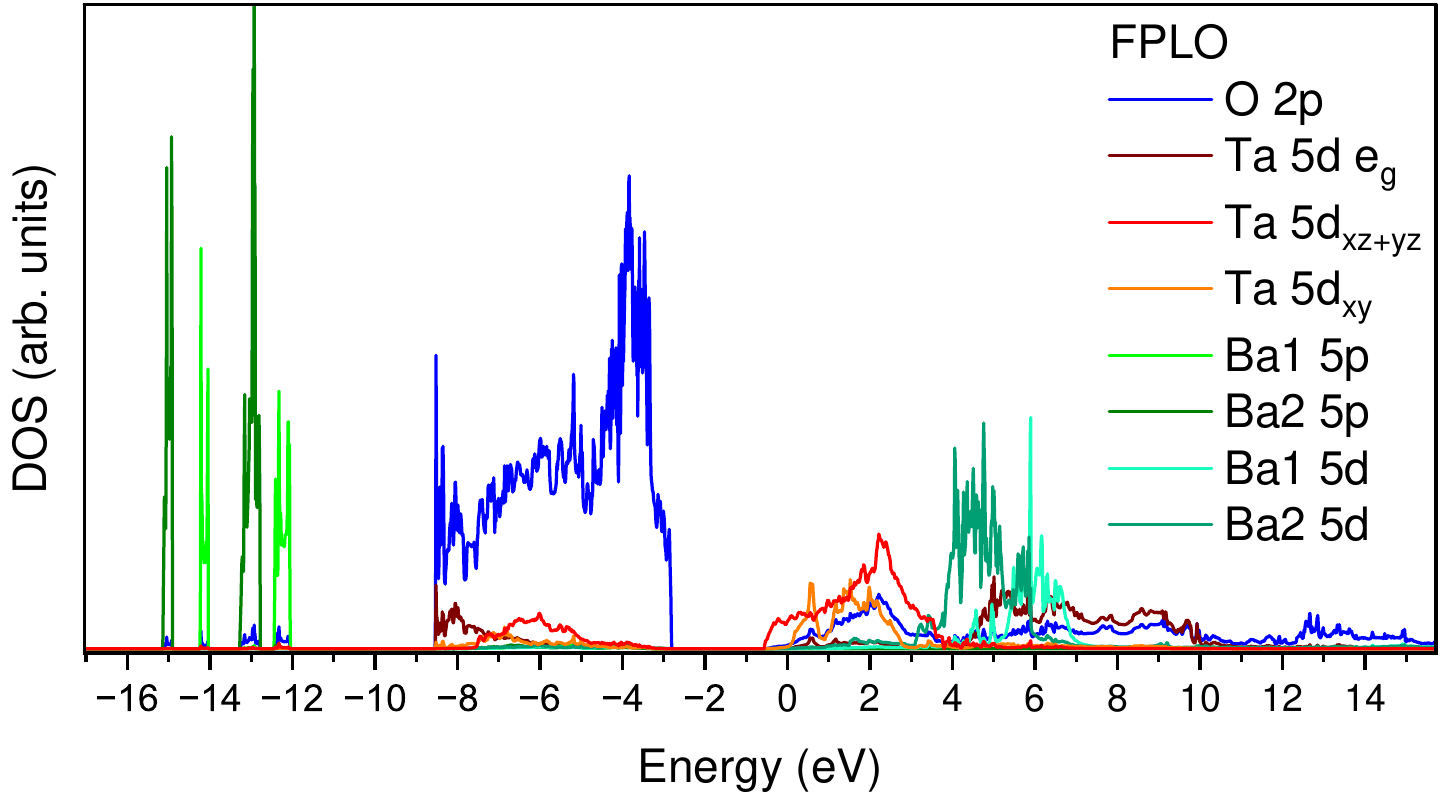}
\end{center}
\caption{PDOS from LDA calculations performed with FPLO for Ba$_3$Ta$_5$O$_{15}$,}
\label{Fig_FPLO}
\end{figure}

\end{document}